\documentclass[]{spie}  

\usepackage{amsmath,amsfonts,amssymb}
\usepackage{graphicx}
\usepackage[colorlinks=true, allcolors=blue]{hyperref}

\title{Performance analysis of a Rydberg atom THz radiometer concept}

\author[a]{Richard J. Murchie}
\author[a]{Petronilo Martin-Iglesias}
\author[a]{Julian Hofmann}
\author[a]{Christophe Buisset}
\affil[a]{ESTEC, European Space Agency, Keplerlaan 1, 2201 AZ, Noordwijk, The Netherlands.}

\authorinfo{Corresponding author: r.jamesmurchie@gmail.com}

\begin{document} 
\maketitle
\begin{abstract}
THz receiver technology lies between electronics and photonics; as such, its implementation is not as mature. This warrants investigation into alternative THz receivers such as Rydberg atom sensors, which are highly susceptible to electric fields. In particular, the conversion architecture has been demonstrated to coherently transduce thermal radiation. This system offers room-temperature operation and does not necessarily require a local oscillator at the frequency of the input signal. We provide an assessment of the feasibility of such sensors for THz radiometry within the context of Earth observation.
\end{abstract}

\section{INTRODUCTION}
\label{sec:intro} 
The mesosphere to lower thermosphere (MLT) is one of the least characterised regions of the Earth’s atmosphere and atomic oxygen is one of the main constituent species. Atomic oxygen does not have convenient optical transitions. Instead, there are two main fine-structure emission lines: one at 2.06~THz and the other at 4.7~THz. In this paper we will focus on sensing the 4.7 THz emission line of atomic oxygen, because of the difficulty of conventional radiometry at this frequency \cite{Richter20}. There is no global and long-term dataset of this atomic species. Therefore, advancements in remote sensing technology suited to probing this atomic species are essential for improving our understanding of the upper atmosphere and the Earth’s climate system.  For example, Keystone, an Earth Explorer 12 candidate mission of the ESA Earth Explorer programme, involves a THz limb sounder which aims to address the gap in our understanding of the MLT \cite{Wright24}. However, in our work, we analyse an alternative THz sensor concept. The purpose of this is to bypass challenges often faced by conventional THz sensor architectures, such as cryogenic cooling. Here we analyse the feasibility and performance of a THz line-selective radiometer concept based on Rydberg atoms.

Rydberg receivers are a promising class of electric field sensor which has the potential for room temperature operation \cite{Fancher21,Allinson26}. This typically involves atoms in an alkali vapour laser-coupled to a Rydberg state (a state with a high principal quantum number): which endows the atomic system with an exaggerated electric dipole moment and thus sensitivity to external electric fields. We focus on the conversion architecture, which involves the coherent transduction of incoming THz radiation to a detected optical field \cite{Kiffner16,Han18,Vogt19,Kumar23,Li24}. This particular architecture has been chosen as there has been experimental demonstration of a Rydberg conversion architecture detecting transduced (microwave) thermal radiation with a room temperature vapour cell\cite{Borowka24}. This study evaluates performance using metrics relevant to radiometry and benchmarks this concept against conventional sensors. We also discuss technological feasibility considerations for the supporting sub-systems. 
\section{Theoretical model}
\subsection{Overview and energy level structure}
A Rydberg receiver involves a vapour cell of atoms. Atomic states are referred to as Rydberg states when they are highly-excited to a large principal quantum number $n$, typically $n>15$. An atom in a Rydberg state exhibits advantageous properties, such as a relatively large lifetime and a high polarisability (sensitivity to electric fields). In particular, we focus on an enriched vapour cell of Rubidium-87 atoms. Moreover, we consider a warm vapour cell with temperature $T=294.15$~K, which is favourable for compact systems and is capable of continuous operation (enabled by the effective repumping of ground state atoms by the Maxwellian velocity of the atoms) \cite{Borowka24}. However, a warm vapour cell means that we have to contend with Doppler linewidth broadening, which can degrade performance of the system. 

An external AC electric field can drive an atomic state transition according to dipole selection rules, where we ignore multipole transitions. This coupling is semi-classically described by an interaction Hamiltonian $\hat{H}(t)_\mathrm{int}=-\hat{\mathbf{d}}\cdot \mathbf{E}(t)$, where $\hat{\mathbf{d}}$ is the electric-dipole operator and $\mathbf{E}(t)$ is the electric-field \cite{Loudon00}. The strength of a dipole transition from an initial state $\vert i\rangle$ to a final state $\vert j\rangle$ is characterised by the dipole transition matrix element \begin{equation}
    \mathbf{d}_{ji}=\langle j \vert \hat{\mathbf{d}} \vert i\rangle.
\end{equation}Semi-classically, an external electric field perturbs the atomic state and can induce coherences between different energy levels, where the expectation value of the dipole-operator $\langle \hat{\mathbf{d}}\rangle$ relates to the single-atom macroscopic polarisation as \begin{equation}
    \mathbf{P}=\mathcal{N}\langle \hat{\mathbf{d}}\rangle=\mathcal{N}\sum_{ij}\rho^\mathrm{lab}_{ij}\mathbf{d}_{ji}, \label{eq:macro_polarization}
\end{equation} where $\mathcal{N}$ is the number density of interacting atoms and $\rho^\mathrm{lab}_{ij}$ is the $(i,j)$ component of the atomic density matrix in the laboratory frame. Equation~\ref{eq:macro_polarization} demonstrates the connection between the density matrix of the atom and the intrinsic atom-field coupling strength. 

The energy difference between Rydberg states with different $n$ principal quantum numbers scales with $\Delta E\propto n^{-3}$. This fact demonstrates that supra-THz radiation is at the upper-end of frequencies accessible to Rydberg-Rydberg transitions. Furthermore, as atomic energy levels are discrete this means that there may not be a Rydberg-Rydberg transition intrinsically near-resonant with a signal of interest. To address this we can apply a static electric field to tune our atomic energy levels (via the Stark effect), such that there is a dipole transition which is now closer to the signal frequency. The Stark effect applied to our Rydberg states can be written as a Hamiltonian, where we assume the static field is along the z-axis \begin{equation} \mathcal{H}_\mathrm{static}=\mathcal{H}_\mathrm{atom}+eE_\mathrm{DC}\hat{z}, \end{equation} where $\mathcal{H}_\mathrm{atom}$ is the Hamiltonian of all the possible energy states in the fine-structure basis and $E_\mathrm{DC}$ is the static electric field in SI units. The static field can cause mixing between the (bare) atomic states and by diagonalising $\mathcal{H}_\mathrm{static}$ we can find the Stark-mixed (dressed) atomic states along with its associated energy level. These dressed states are linear combinations of the bare atomic states. For low $n$ atomic states, where we use the hyperfine-structure basis, under the assumption the static field is not too strong, this Stark effect approximately manifests as an energy shift $\Delta E_{\mathrm{stark}}$ of the bare states. We also define a quantisation axis, by placing a static magnetic field on the z-axis of our sensor: which enables a driving field's polarisation to cleanly determine a dipole transition based on the change of the magnetic quantum number.

As there is an infinite number of atomic states, for computational tractability we want to constrain our system to only consider atomic states that strongly couple with our fields. Hence, we consider six bare atomic states, labelled by Dirac notation as $\vert 1\rangle,...,\vert 6\rangle$. This reduction of the Hilbert space of the atomic Hamiltonian is possible as we utilise a phase-matched, closed-loop six-wave-mixing manifold driven through maximally polarized states (apart from $\vert 6\rangle$). This manifold also has the benefit of suppressing dark state accumulation. Even with this manifold choice there are many spontaneous decay and blackbody radiation (BBR)-induced transition channels. For the sake of simplicity spontaneous decay channels that are not intra-manifold are grouped into a loss reservoir state $\vert\mathrm{loss}\rangle$. We also involve BBR-induced transition channels and follow the same criteria if they are not intra-manifold.

The bare state manifold we select is as follows (in the hyperfine structure basis for $\vert 1\rangle,\vert 2\rangle,\vert 6\rangle$ and fine structure basis for $\vert 3\rangle,\vert 4\rangle,\vert 5\rangle$) 
\begin{align*} \vert 1\rangle&:=\vert 5S_{1/2},F=2,m_F=2\rangle, \\ \vert 2\rangle&:=\vert 5P_{3/2},F=3,m_F=3\rangle, \\ \vert 3\rangle&:=\vert 22D_{5/2},m_I=3/2,m_J=5/2\rangle, \\ \vert 4\rangle &:= \vert 36 P_{3/2},m_I=3/2,m_J=3/2\rangle, \\ \vert 5\rangle &:= \vert 36S_{1/2},m_I=3/2,m_J=1/2\rangle, \\ \vert 6\rangle &:=\vert 5P_{1/2},F=2,m_F=1\rangle. 
\end{align*} 

The Stark effect and Zeeman effect cause state mixing, which, as mentioned earlier, can be approximated as an energy shift/detuning on the bare state for $\vert 1\rangle$, $\vert 2\rangle$, and $\vert 6\rangle$. We define the dressed state associated to the initial bare states as $\vert \tilde{3}\rangle$, $\vert \tilde{4}\rangle$, and $\vert \tilde{5}\rangle$. We do not provide the exact linear combination of atomic states for each dressed state; however, we ensure that mixing is low enough such that the initial bare state is retained as the primary component. Therefore, our manifold is \begin{equation}\mathcal{M}=\{\vert 1\rangle,\vert 2\rangle,\vert\tilde{3}\rangle,\vert\tilde{4}\rangle,\vert\tilde{5}\rangle,\vert 6\rangle\}.\end{equation}

\begin{figure}
        \centering
        \includegraphics[width=\linewidth]{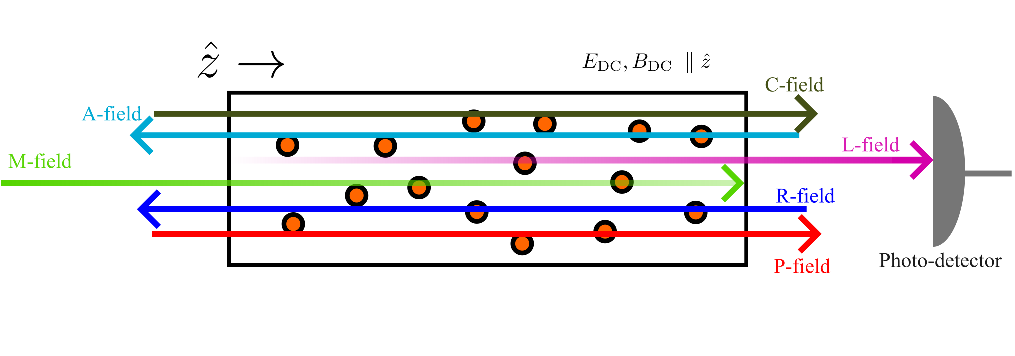}
        \caption{Schematic of Rydberg vapour cell and the interaction of the six fields, with their propagation directions. The M-field as the incoming signal of interest field and the L-field as the generated field which we detect. The z-axis is shown to be parallel to the collinear fields and we state that the static electric and magnetic fields are parallel to the z-axis.}
        \label{fig:Rydberg_schematic}
    \end{figure}

Associated to each step-wise atomic transition in our manifold is a field. This field may be an auxiliary/pump laser, it may be the signal of interest or it may be a generated field. We define the fields (with a field label $x$) according to their positive frequency component as\begin{equation}
\mathbf{E}^+_x(\mathbf{r},t)=\mathbf{e}_x\mathcal{E}_x e^{i(\mathbf{k}_x\cdot\mathbf{r}-\omega_x t)},
\end{equation} where $\mathbf{e}_x$, $\mathcal{E}_x$, $\mathbf{k}_x$ and $\omega_x$ are the unit polarisation vector, envelope function, wave-vector and angular frequency for the field, respectively. From this we define the Rabi frequency as
\begin{equation} \Omega_x=\frac{\mathbf{d}_{ji}\cdot \mathbf{e}_x}{\hbar}\mathcal{E}_x, \end{equation} where $\mathbf{d}_{ji}$ is the dipole transition matrix element for state $i$ to state $j$. We use the convention that each field is defined by its absorptive pathway. The pump laser (P) drives the absorptive leg $\vert 1\rangle\to \vert 2\rangle$ with Rabi frequency $\Omega_\mathrm{P}$. The coupling laser (R) drives the absorptive leg $\vert 2\rangle \to\vert \tilde{3}\rangle$ with Rabi frequency $\Omega_\mathrm{R}$. The incoming THz radiation (signal of interest) drives the absorptive leg $\vert \tilde{3}\rangle$ to $\vert \tilde{4}\rangle$ with Rabi frequency $\Omega_\mathrm{M}$. The decoupling local oscillator (C) drives the stimulated emission leg $\vert \tilde{4}\rangle$ to $\vert \tilde{5}\rangle$ with Rabi frequency $\Omega^*_\mathrm{C}$. The auxiliary laser (A) drives the stimulated emission leg $\vert \tilde{5}\rangle$ to $\vert 6\rangle$ with Rabi frequency $\Omega^*_A$. Lastly, the generated optical field (L) drives the transition leg $\vert 6\rangle $ to $\vert 1\rangle$ with Rabi frequency $\Omega^*_L$. The overlap region of all our collinear fields is what specifies the interaction region of the atom-light system with width $r_\mathrm{int}=1.5$~mm. We make the simplifying assumption, that in the paraxial regime, the (relatively strong) $P,R,C,A$ control field envelopes are spatially homogeneous over the interaction region, which requires that these fields are wider than the $M$ and $L$ fields. Hence, the control field power requirements are higher than what is inferred from their Rabi frequencies. Figure~\ref{fig:Rydberg_schematic} is a simple schematic of our Rydberg conversion scheme. 

    We define an atomic-state referenced frequency with its manifold labels as \begin{equation}
    \omega_{ij}=\frac{(E_i-E_j)e}{\hbar} \;[\mathrm{rad/s}],
\end{equation} where $E_{i/j}$ are in state energies in electron volts and $e$ is the electron charge constant. By contrast, we specify our fields according to their label $\omega_{\mathrm{P/R/M/C/A/L}}$. These fields can be detuned from exact resonance with their associated atomic transitions. 

\begin{figure}
        \centering
        \includegraphics[width=0.5\linewidth]{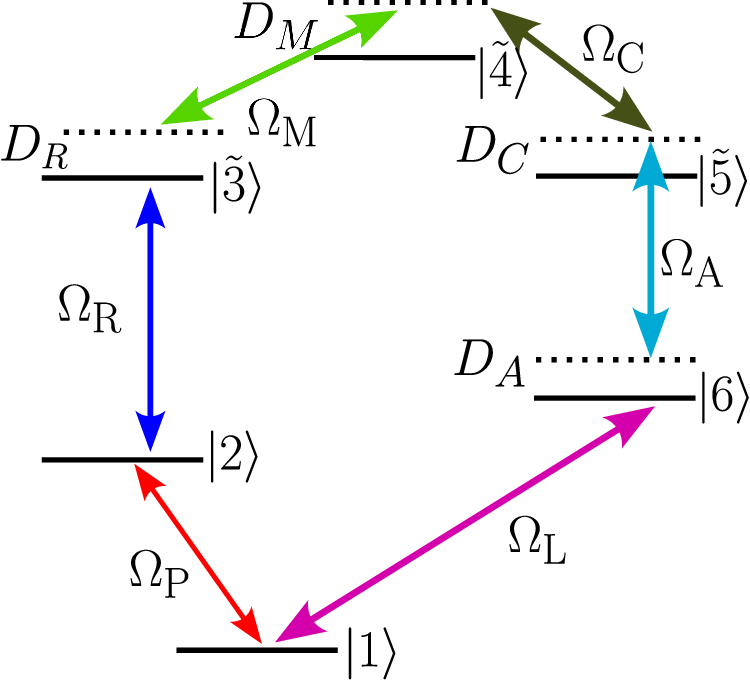}
        \caption{Energy level diagram of the Stark-dressed 6-level closed loop manifold.}
        \label{fig:energy_diagram}
    \end{figure}

This system leverages loop-closure to select a coherent six-wave-mixing pathway from the manifold \cite{Kiffner16,Han18,Vogt19}. From this condition, the absorbed and emitted photons satisfy the required angular-momentum $\Delta q=0$ and energy balances $\Delta \omega=0$ such that the induced polarisation has a component which oscillates with the frequency of the generated L-field. The phase-matching conditions for coherent propagation are also determined from loop-closure. We have chosen a manifold such that the angular momentum of each field $q_x$ satisfies $\Delta q=q_p+q_R+q_M-q_C-q_A-q_L=0$. Furthermore, due to loop-closure the generated L-field has an angular frequency that linearly corresponds to the angular frequency of the incoming signal (M-field) $\omega_L=\omega_P+\omega_R+\omega_M-\omega_C-\omega_A$. Figure~\ref{fig:energy_diagram} visualises the closed-loop nature of this manifold of Stark-dressed transitions coupled by our fields, with field-referenced detunings $D_{R/M/C/A}$. 
    
Table~\ref{table:transition_properties} details some key attributes associated to each field, for the set static electric field and static magnetic field. The atomic physics Python package $\textit{ARC}$ is used to calculate the atomic transitions and dipole matrix elements for our system \cite{Weatherill17}. The polarisation of the fields, from the convention of $+z$ propagation, is either right-hand circular polarised ($\sigma^+$) or left-handed circular polarised ($\sigma^-$), this is because we want our fields to propagate parallel to the quantisation axis $\hat{z}$. We have that $s_x=\pm 1$ denotes the propagation direction $(+z/-z)$ of our field. Hence, the polarisation requirement of a field is easily obtained from assessment of the field angular momentum $q_x$ and its propagation direction $s_x$. Due to the closed-loop condition it is apparent that the generated L-field (from the $\vert6\rangle \to\vert 1\rangle$ coherence) is left-hand circularly polarised.
\begin{table}[htbp]
\centering
\begin{tabular}{cccccc} 
 \hline
 Field & Angular momentum & Atomic transition & Dipole element & Direction & Rabi frequency \\ [0.5ex] 
 \hline
 P & $q_p=1$ & $780.246$~nm & $d_{21}=2.98915$ & $s_P=1$ & $10$ \\ 
 R & $q_R=1$ & $485.036$~nm & $d_{\tilde{3}2}=-5.7367\times10^{-2}$ & $s_R=-1$ & 20 \\
 M & $q_M=-1$ & $4.7447716$~THz & $d_{\tilde{4}\tilde{3}}=2.04772$ & $s_M=1$ &  \\
 C & $q_C=1$ & $80.81$~GHz & $d_{\tilde{4}\tilde{5}}=670.057$ & $s_C=1$ & 60 \\
 A & $q_A=1$ & $475.959$~nm & $d_{\tilde{5}6}=-4.34\times10^{-3}$ & $s_A=-1$ & 1.5 \\ 
 L & $q=-1$ & $794.98$~nm & $d_{61}=-1.221$ & $s_L=1$ &  \\ 
 \hline
\end{tabular}
\caption{Basic properties associated to each field and transition. The field may be detuned from the associated atomic transition. The dipole matrix elements are scalars as we have a specified polarisation and they are in atomic units (a.u.). The static DC field for Stark dressing on the z-axis is $E_z=2935.5$~V/m and the static bias magnetic field also on the z-axis is $B_z=10$~G. The Rabi frequencies are in units of Mrad/s. As both the L- and M- fields are not control fields, we do not specify a Rabi frequency.}
\label{table:transition_properties}
\end{table}

\subsection{Hamiltonian}
The Gorini-Kossakowski-Sudarshan-Lindblad (GKSL) master equation describes the Markovian time evolution of an open quantum system, here we make the dipole and rotating wave approximations \cite{Scully97}. In particular, we wish to track the atomic density operator $\hat{\rho}$. This refers to the state of a single atom, which is found by tracing out the Hilbert spaces of the involved light fields out of the total density matrix of the atom-light system. It has the form, \begin{equation}
    \dot{\hat{\rho}}=-\frac{i}{\hbar}\left[\hat{H},\hat{\rho}\right]+\sum_\lambda\mathcal{L}^\lambda_\mathrm{decay}\hat{\rho}+\sum_\lambda\mathcal{L}^\lambda_\mathrm{deph}\hat{\rho},\label{eqn:master_eqn}
\end{equation} where $\hat{H}$ is the Hamiltonian and the Lindbladian dissipators for decay dynamics $\mathcal{L}^\lambda_\mathrm{decay}$ (population transfer and damping of coherences) and dephasing dynamics $\mathcal{L}^\lambda_\mathrm{deph}$ (no population transfer), for an underlying physical cause as labelled by $\lambda$. 

Our Hamiltonian of the semi-classical atom-light system in the basis of our Stark/Zeeman shifted and dressed state manifold ($\mathcal{M}$) is \begin{equation}
    \hat{H}=-\hbar\sum^6_{j=2}\delta^\mathrm{eff}_j\vert j\rangle\langle j\vert-\hbar\left(\Omega_\mathrm{P}\vert 2\rangle\langle 1\vert+\Omega_\mathrm{R}\vert \tilde{3}\rangle\langle 2\vert+\Omega_\mathrm{M}\vert \tilde{4}\rangle\langle \tilde{3}\vert+\Omega_\mathrm{C}\vert \tilde{4}\rangle\langle \tilde{5}\vert+\Omega_\mathrm{A}\vert \tilde{5}\rangle\langle 6\vert+\Omega_\mathrm{L}\vert 6\rangle\langle 1\vert+\mathrm{H.c} \right),
\end{equation} where we have applied the dipole and rotating wave approximation, along with the use of a rotating frame to remove fast oscillating terms. The cumulative (effective) detuning is $\delta^\mathrm{eff}_j$ incorporates all the effects that we consider that contribute to each state's cumulative detuning. Our Hamiltonian is semi-classical as our fields are classical (later in this paper the signal (M) and generated optical (L) field will be quantised, whereas the other fields are strong and thus can remain approximately classical). The effects that we explicitly model are: spontaneous decay, BBR transitions \cite{Beterov09}, laser phase noise, (elastic) collisional broadening \cite{Loudon00}, dipole-mediated atom-atom effects \cite{Kiffner16}, transit broadening, and collisional-ionisation into a ionisation-loss reservoir state $\vert \mathrm{ion}\rangle$. As previously mentioned, we involve a loss reservoir state $\vert \mathrm{loss}\rangle$ to account for non-intra-manifold spontaneous decay and BBR transitions. Lastly, we account for Doppler broadening, which affects the effective detuning for each manifold state. Non-exhaustively, we neglect BBR-induced detuning, many-body effects, cooperative effects, and ionisation-based detuning \cite{Vitali11,2Beterov09,Prokop26}.

\subsubsection{Detuning terms}
Each manifold state $\vert j\rangle$ has a cumulative effective detuning term $\delta^\mathrm{eff}_{j}$ which comprises a cumulative rotating frame frequency, the energy from ground state, a Doppler detuning from a particular velocity class, and an atom-atom detuning. As discussed earlier,  Instructed by the rotating frame frequencies, we define the cumulative wave-vectors and cumulative effective detuning (for a velocity class $v$) for each manifold state ($\vert 1\rangle,\dots,\vert 6\rangle)$ in Table~\ref{tab:cumulative_detunings}.
\begin{table}[t]
\centering
\small
\begin{tabular}{@{}c@{\qquad}c@{}}
$
\begin{aligned}
K_1 &= 0,\\
K_2 &= s_P k_P,\\
K_3 &= K_2+s_R k_R,\\
K_4 &= K_3+s_M k_M,\\
K_5 &= K_4-s_C k_C,\\
K_6 &= K_5-s_A k_A,
\end{aligned}
$
&
$
\begin{aligned}
\delta^\mathrm{eff}_1(v) &= 0,\\
\delta^\mathrm{eff}_2(v) &= \omega_P-\omega_{21}-K_2v,\\
\delta^\mathrm{eff}_3(v) &= \omega_P+\omega_R-\omega_{31}-K_3v,\\
\delta^\mathrm{eff}_4(v) &= \omega_P+\omega_R+\omega_M-\omega_{41}-K_4v,\\
\delta^\mathrm{eff}_5(v) &= \omega_P+\omega_R+\omega_M-\omega_C-\omega_{51}-K_5v,\\
\delta^\mathrm{eff}_6(v) &= \omega_L-\omega_{61}-K_6v.
\end{aligned}
$
\end{tabular}
\caption{Cumulative wavevectors and velocity-dependent effective detunings, with $k_x=\omega_x/c$ for the collinear geometry.}
\label{tab:cumulative_detunings}
\end{table}

We can also define edge (field-referenced) detunings, which indicate how detuned each field is from its associated atomic transition. As such, dropping the Doppler velocity class argument, for our particular system we have the values (in Mrad/s) $D_P=\delta^\mathrm{eff}_2-\delta^\mathrm{eff}_1=0$, $D_R=\delta^\mathrm{eff}_3-\delta^\mathrm{eff}_2=10$, $D_M=\delta^\mathrm{eff}_4-\delta^\mathrm{eff}_3$, $D_C=\delta^\mathrm{eff}_4-\delta^\mathrm{eff}_5=-7.241$, $D_A=\delta^\mathrm{eff}_5-\delta^\mathrm{eff}_6=38.9655$, and $D_L=\delta^\mathrm{eff}_6-\delta^\mathrm{eff}_1$. Both the M- and L-field do not have edge detunings specified as these are our dynamic fields, whereas the control fields have a static edge detuning.
\subsection{Atomic response}
\label{sect:atom_response}
We acquire the steady state solution $\hat{\rho}_\mathrm{ss}$ of our atomic system when the time-derivative in Eq.~\ref{eqn:master_eqn} equals zero ($\dot{\hat{\rho}}=0$). As we consider a warm vapour cell \cite{Borowka24}, we must average the steady state solution over Doppler velocity classes, weighted by the Maxwellian velocity profile \begin{equation}
    \bar{\hat{\rho}}_\mathrm{ss}=\int dv \sqrt{\frac{m_{87}}{2\pi k_B T}}e^{-\frac{m_{87} v^2}{ 2 k_B T}}\hat{\rho}_\mathrm{ss}(v).
\end{equation} This can be numerically performed in the Rydberg physics Python package \textit{Rydiqule} \cite{Miller24}, or by a recently developed analytic technique for Doppler averaging \cite{Nagib25}. The temperature of the cell is important as it affects the vigour of the Doppler broadening and it affects the density of the vapour cell (which in turn affects the density of Rydberg atoms in the interaction region).

\begin{figure}
    \centering
    \includegraphics[width=\linewidth]{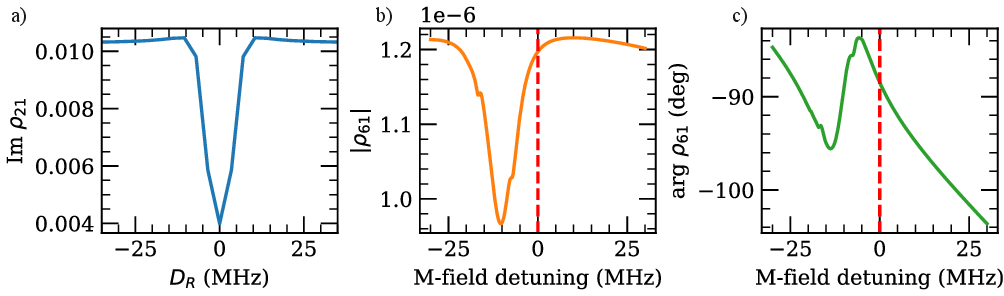}
    \caption{a) 3-level EIT probe field absorption dip, with 3-level EIT FWHM linewidth $\gamma_\mathrm{EIT}=9.25$~MHz over R-field detuning from atomic transition frequency $\omega_{32}$. The atomic coherence $\rho_{61}$ is scanned over M-field detuning with respect to the atomic oxygen transition for b) magnitude $\vert \rho_{61}\vert$ and c) argument $\mathrm{arg}(\rho_{61})$. The vertical dashed red line denotes the atomic oxygen ($4.7$~THz) central frequency.}
    \label{fig:atomic_response}
\end{figure}

As we have an expression to calculate the atomic response, we can find the zeroth-order Doppler-averaged solution (when $\Omega_M=0$ and $\Omega_L=0$) of the Rydberg $\vert \tilde{3}\rangle$ state population $\bar{\hat{\rho}}^{(0)}_{33}$ \cite{Finkelstein23}. This solution is approximately equivalent to a 3-level EIT system. We operate in a regime such that the M-field is weak, this ensures that $\vert\tilde{3}\rangle$ is the dominant contribution to the Rydberg atom population $N_{ryd,pop}\propto \bar{\hat{\rho}}^{(0)}_{33}$. Following this, we can characterise the 3-level EIT full-width half-maximum (FWHM) linewidth $\gamma_\mathrm{EIT}$, as shown in Fig.~\ref{fig:atomic_response}. Our numerical analysis using atomic data-sheets and the zeroth-order Doppler-averaged solution indicates that the Rydberg blockade effect (from Van der Waals detuning) is not problematic for our operating regime \cite{Steck25}.

In the regime where our dynamic fields $\Omega_M$ and $\Omega_L$ are weak, we linearise the atomic coherence terms pertaining to the dynamic fields as follows \begin{align}
    \rho_{61} &\approx \xi^M_{61}\Omega_M+\xi^L_{61}\Omega_L, \\
    \rho_{43} &\approx \xi^M_{43}\Omega_M+\xi^L_{43}\Omega_L.
\end{align} The cross-term coefficients $\xi^M_{61}$ and $\xi^L_{43}$  determine how the M or L field generates a coherence on the $\vert 6\rangle\to \vert 1\rangle$ or $\vert 3\rangle\to\vert 4\rangle$ transitions, respectively \cite{Kiffner16}. Conversely, the coefficients $\xi^M_{43}$ and $\xi^L_{61}$ correspond to self-response of the M/L fields on their respective atomic transitions. These self-response terms demonstrate the imperfection of this system as a frequency converter. We invoke a weak coherent test M-field and L-field to estimate the cross-term and self-response atomic coefficients. This numerical procedure is necessary, as for a stationary thermal input the mean field, and hence the ensemble-averaged generated-field amplitude, vanishes. The thermal signal instead resides in the second-order field correlation and output photon occupation \cite{Kaur26}. Figure~\ref{fig:atomic_response} plots the 3-level EIT linewidth and both the magnitude and argument of the atomic coherence $\rho_{61}$, for coherent test Rabi frequencies $\Omega_L=\Omega_M=1$~krad/s. The validity of the linear regime has been checked over a suitable range of coherent test Rabi frequencies $\Omega_{M/L}$.

\subsection{Propagation}
The nonlinear wave equation is \begin{equation}
    \left(\frac{1}{c^2}\frac{\partial^2}{\partial t^2}-\nabla^2\right)\mathbf{E}=-\frac{1}{c^2\epsilon_0}\frac{\partial^2}{\partial t^2}\mathbf{P},\label{eq:NL_waveeqn}
\end{equation} where $\mathbf{E}=\mathbf{E}_L+\mathbf{E}_M$ and the single-atom macroscopic polarisation is $\mathbf{P}$ \cite{Boyd08}. In our specified geometry we have a non-zero phase-mismatch $\Delta k=s_L k_L-(s_P k_P+s_R k_R+s_M k_M-s_C k_C-s_A k_A)\neq 0$, as we employ counter-propagating P and R fields to reduce Doppler broadening. However, we take the zero phase-mismatch limit. This approximation can in principle be reached as a non-collinear geometry could facilitate $\Delta k\approx0$, at the expense of imperfect projection of the beam polarisation onto components of the atomic dipole transitions, with respect to the common atomic quantisation axis. Practical considerations remain, and we constrain the length of the cell such that for a reasonably small tilt from a pair of lasers, phase coherence is facilitated over the cell length of $L_\mathrm{cell}=1$~cm. For analytic simplicity, our model remains in the collinear-reduction for treating the wave-vectors.

The single-atom positive-frequency macroscopic polarisation (neglecting atom-atom interactions and in the zero phase-mismatch limit) pertaining to our L- and M-fields can be written as
\begin{equation}
\mathbf{P}^+_M+\mathbf{P}^+_L=\mathcal{N}\left(\mathbf{d}_{34}\rho_{43}e^{i(k_M z-\omega_M t)}+\mathbf{d}_{16}\rho_{61}e^{i(k_L z-\omega_L t)}\right).
\end{equation} From $\mathbf{E}^+_M$ and $\mathbf{E}^+_L$ we have two independent wave equations for their respective positive-frequency polarisation terms $\mathbf{P}^+_M\propto \rho_{43}$ and $\mathbf{P}^+_L\propto \rho_{61}$. The solution of the nonlinear wave equation (Eq.~\ref{eq:NL_waveeqn}) is derived in Appendix~\ref{appendix:transfer_function}, where we acquire a transfer function $H(\omega,z)$ which determines how an input M- and L-field results in the output M- and L-field, for each spectral component after cell length $z$. In effect, we have a two-port lossy frequency converter, described by the relevant components of the transfer function: for M$\to$L conversion $H_{LM}$ and for L$\to$M conversion $H_{ML}$. Of particular interest is ($\mathrm{M}\to\mathrm{L}$) conversion, which we explicitly define as \begin{equation}
    H_{LM}(\omega,z)=\frac{ie^{\frac{i}{2}(\eta_M(\omega) \xi^M_{43}(\omega)+\eta_L(\omega_L)\xi^L_{61}(\omega_L))z}\eta_L(\omega_L)\xi^M_{61}(\omega_L)}{K(\omega)}\mathrm{sin}(K(\omega)z),
\end{equation} where \begin{equation}
        K(\omega)=\sqrt{\left(\frac{\eta_M(\omega) \xi^M_{43}(\omega)-\eta_L(\omega_L)\xi^L_{61}(\omega_L)}{2}\right)^2+\eta_M(\omega) \xi^L_{43}(\omega)\eta_L(\omega_L)\xi^M_{61}(\omega_L)}.
\end{equation}

\subsection{Quantisation}
Our signal may be weak enough such that quantisation of the dynamic fields (M and L) is required. Moreover, our control fields are strong enough such that we can assume that there is no pump-depletion, otherwise a Manley-Rowe treatment would be required \cite{Boyd08}. The linear regime for our dynamic fields ensures straightforward translation from classical field amplitudes to quantum field operators. We assume that the quantised M- and L- fields have matched transverse spatial modes. Whereas the strong control fields are transversally-broad such that their Rabi frequencies, and hence the linearised atomic coherences, are approximately homogeneous across the interaction region.

We introduce a travelling continuum mode operator $\hat{a}_x$, with commutator \begin{equation}
    [\hat{a}_x(\omega),\hat{a}_y^\dag(\omega^{'})]=2\pi \delta_{xy}\delta(\omega-\omega^{'}),
\end{equation} where we have used the Kronecker delta and the Dirac delta function.
For a stationary field its first-order frequency correlation is \begin{equation}
    \langle \hat{a}^\dag(\omega)\hat{a}(\omega^{'})\rangle=2\pi n(\omega)\delta(\omega-\omega^{'}),\label{eq:freq_corr}
\end{equation} with $n(\omega)$ as the spectral mode occupation. For a travelling paraxial mode $x$ with effective area $A_x$ we define its field envelope as\begin{equation}
    \hat{\mathcal{E}}_x(\omega)=\sqrt{\frac{\hbar\omega}{2\epsilon_0 c A_x}}\hat{a}_x(\omega),
\end{equation} which has been chosen to ensure photon-flux normalisation, as detailed in Appendix~\ref{appendix:normalisation}. In the ideal/lossless limit this system implements a coherent beamsplitter-like interaction between the M-field and L-field, for which the total photon flux is conserved.

\begin{figure}
    \centering
    \includegraphics[width=\linewidth]{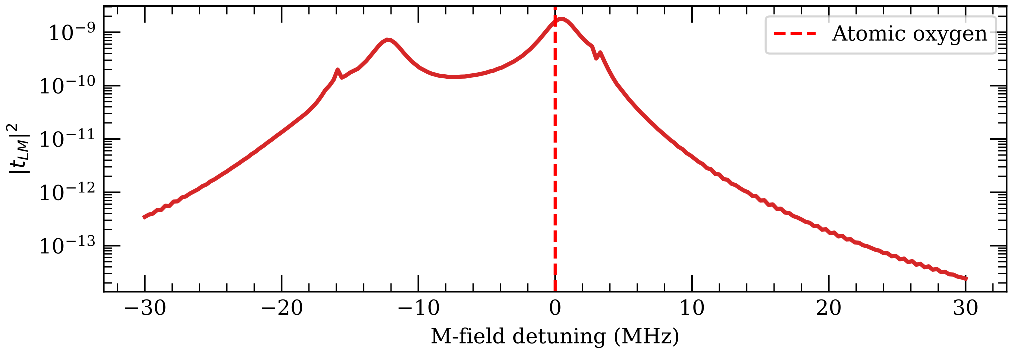}
    \caption{The $\mathrm{M}\to\mathrm{L}$ conversion function magnitude $\vert t_{LM}\vert^2$ plotted with respect to M-field detuning from the atomic oxygen emission centre frequency.}
    \label{fig:conversion_response}
\end{figure}

We recast our input-output Rabi frequency equation (Eq.~\ref{eq:input_output_rabi}) into the more suitable photon-flux normalised operator form for assessing conversion efficiency and output photon statistics. Thus, a Rabi frequency for a quantised field is defined as $\hat{\Omega}_x(\omega)=g_x(\omega)\hat{a}_x(\omega)$, where $g_x(\omega)=\frac{\mathbf{d}_{ji}\cdot \mathbf{e}_x}{\hbar}\sqrt{\frac{\hbar \omega}{2\epsilon_0 c A_x}}$. From this quantised definition of the Rabi frequency and introduction of the Langevin noise operators $\hat{F}_M(\omega)$ and $\hat{F}_L(\omega)$ \cite{Barnett02}, we can now recast the solution of the nonlinear wave equation (Eq.~\ref{eq:input_output_rabi}) as \begin{equation}
    \begin{pmatrix}
        \hat{a}_M(z,\omega) \\ \hat{a}_L(z,\omega_L)
    \end{pmatrix}=S\begin{pmatrix}
        \hat{a}_M(0,\omega) \\ \hat{a}_L(0,\omega_L)
    \end{pmatrix}+\begin{pmatrix}
        \hat{F}_M(\omega) \\ \hat{F}_L(\omega_L)
    \end{pmatrix}=\begin{pmatrix}
       H_{MM} & \frac{g_L(\omega_L)}{g_M(\omega)}H_{ML} \\ \frac{g_M(\omega)}{g_L(\omega_L)}H_{LM} & H_{LL}
    \end{pmatrix}\begin{pmatrix}
        \hat{a}_M(0,\omega) \\ \hat{a}_L(0,\omega_L)
    \end{pmatrix}+\begin{pmatrix}
        \hat{F}_M(\omega) \\ \hat{F}_L(\omega_L)
    \end{pmatrix},
\end{equation} the introduction of the Langevin noise operators is essential to preserve the commutation relation of the output mode operators, as in general $S$ is not unitary ($S S^\dag\neq \hat{1})$. We use a vacuum-noise model for the environmental modes present in the Langevin noise operators, as a lower bound on the noise. A more complete Heisenberg-Langevin treatment would derive the associated reservoir fluctuations, including L-field fluorescence and thermally-driven atomic noise. Hence, our Langevin noise operators are \begin{equation}
    \begin{pmatrix}
        \hat{F}_M(\omega) \\ \hat{F}_L(\omega_L)\end{pmatrix}=\sqrt{\hat{1}-S S^\dag}\begin{pmatrix}
            \hat{v}_M(\omega)\\ \hat{v}_L(\omega_L)
    \end{pmatrix},
\end{equation} where $\hat{v}_{M/L}$ are vacuum environmental modes with canonical commutators for a travelling continuum mode. As M-field to L-field conversion is our main concern we define $t_{LM}=\frac{g_M(\omega)}{g_L(\omega_L)}H_{LM}$ for brevity. Figure~\ref{fig:conversion_response} plots the $\mathrm{M}\to\mathrm{L}$ conversion function magnitude. From comparison to Fig~\ref{fig:atomic_response} it is clear that the maximum atomic coherence magnitude does not necessarily correspond to maximum conversion efficiency. 

\subsection{Photon statistics}
As we have an input-output relation in terms of the bosonic mode operators for the vapour cell, we can proceed with calculating the photon statistics of the output (converted) field. For the cell-length $L_\mathrm{cell}$, our output L-field bosonic mode operator is $\hat{a}_L(z=L_\mathrm{cell},\omega)$. The input M-field operator is decomposed as \begin{equation}
    \hat{a}_M(0,\omega)=\sqrt{\eta_\mathrm{c}}\hat{a}_\mathrm{M,scene}+\sqrt{1-\eta_\mathrm{c}}\hat{a}_\mathrm{M,sensor},
\end{equation} this decomposition represents a simple accepted-mode description where coupling loss from the scene $\hat{a}_\mathrm{M,scene}$ to our sensor interaction region is represented by $\sqrt{\eta_\mathrm{c}}$, from which the complement ($\sqrt{1-\eta_\mathrm{c}}$) involves the contribution from sensor-induced BBR $\hat{a}_\mathrm{M,sensor}$. Since the sensor-induced BBR is treated as an additive background independent of the scene-coupling loss, it has its mode occupation scaled $n_\mathrm{M,sensor}\to \frac{n_\mathrm{M,sensor}}{1-\eta_\mathrm{c}}$ to avoid being affected by the scene loss-channel, for $\eta_\mathrm{c}<1$. Moreover, we assume that the initial M-field, L-field and vacuum environment modes are independent, which simplifies calculation of the photon statistics. A more complete treatment would explicitly determine the angular thermal mode acceptance from the atomic response \cite{Borowka24}. As the input L-field is in the vacuum and the Langevin noise operator corresponds to the vacuum, the first-order frequency correlation of the converted L-field is \begin{equation}
    \langle \hat{a}^\dag_L(z=L_\mathrm{cell},\omega_L)\hat{a}_L(z=L_\mathrm{cell},\omega_L^{'})\rangle=\vert t_{LM}\vert ^2\langle \hat{a}^\dag_M(0,\omega)\hat{a}_M(0,\omega^{'})\rangle,\label{eq:L_freq_corr}
\end{equation}similarly, we have assumed independence between the scene signal and sensor-induced BBR. The scene $n_\mathrm{scene}(\omega)$ and the accepted-mode sensor-induced BBR $n_\mathrm{sensor}(\omega)$ mean photon occupation per spectral mode can be calculated from Planck's law. From Eq.~\ref{eq:freq_corr} and Eq.~\ref{eq:L_freq_corr}, we have the mean photon occupation for the output L-field per spectral mode  \begin{equation}
    n_L(\omega_L)=\vert t_{LM}\vert^2 \left(\eta_\mathrm{c} n_\mathrm{scene}(\omega) +n_
     \mathrm{sensor}(\omega)\right).
\end{equation} A thermal state mixing with the vacuum or another thermal state is still thermal, hence the variance per spectral mode is \begin{equation}
    \mathrm{Var}(n_L(\omega_L))=n_L(\omega_L)(1+n_L(\omega_L)).
\end{equation}

In practice, we record photons over an integration time $\tau_d$ on an imperfect photo-detector with detector efficiency $\eta_d$, which includes the vapour cell output coupling loss and detector quantum efficiency. The ambient background at the L-field is negligible (ignoring fluorescence from the sensor) and so can be approximated as the vacuum. We also consider a moderately cooled low-noise SPAD, which has a dark count rate of $\mathrm{DCR}=10$~Hz. The detected spectral occupation is $n_\mathrm{det}(\omega_L)=\eta_d n_L(\omega_L)$. Consequently, the mean detected photon count over integration time $\tau_d$ is \begin{equation}
    \langle N_L\rangle=\tau_d \int_\mathcal{B}\frac{d\omega_L}{2\pi} n_\mathrm{det}(\omega_L)+\tau\mathrm{DCR},
\end{equation} where the integration is taken over the bandwidth $\mathcal{B}$ of concern. Furthermore, the photon-count variance is \begin{equation}
    \mathrm{Var}(N_L)=\tau_d \int_\mathcal{B}\frac{d\omega_L}{2\pi} n_\mathrm{det}(\omega_L)\left( 1+n_\mathrm{det}(\omega_L)\right)+\tau\mathrm{DCR}.~\label{eq:photocount_var}
\end{equation}
\subsection{Performance}
The optical read-out is subject to several intrinsic and technical noise sources which affect the performance of our system. Quantum projection noise (QPN) is associated to the standard quantum limit \cite{Chen26}. It arises from measurement determining the probabilistic state of an atom, for an ensemble of $N$ atoms this amounts to $N$ random processes, which then exhibits a statistical spread. Without inducing quantum correlations, QPN represents a lower limit on the performance of the system and is negligible compared to the other noise sources present.  A major limitation of conversion systems is photon shot noise, which arises due to the discrete nature of photo-counting. Our thermal signal has both a Poissonian component and a thermal bunching component: which is apparent in the photo-count variance in Eq.~\ref{eq:photocount_var}. Sensor-induced BBR at the M-field frequency band enters through a thermal-loss channel and is propagated through the same conversion channel as our signal of interest, it adds to the noise floor of our system. Furthermore, the conversion response is dependent on the lasers, hence any laser noise can cause unwanted variation in the response. In particular, laser phase (frequency) noise is modelled with a dephasing term in the Master equation. This noise source arises as we do not consider a phase-locked system and our lasers have a finite linewidth. Relative-intensity noise and laser frequency drift would produce fluctuations/drift in the atomic conversion efficiency and consequently the number of converted photons; however, we do not consider it in our model. Lastly, we consider detector dark counts as a Poissonian electronic noise source. 

As we are sensing thermal radiation a natural metric for performance is noise equivalent power (NEP). In particular, we define the input-referred intrinsic receiver $\mathrm{NEP}_\mathrm{sensor}$ as \begin{align}
    \mathrm{NEP}_\mathrm{sensor}&=\frac{\sqrt{2\tau_d\mathrm{Var}(N_L(n_\mathrm{scene}(\omega)=0))}}{\vert\frac{\partial\langle N_L (n_\mathrm{scene}(\omega)=0)\rangle}{\partial P_\mathrm{sensor}}\vert} \;[\mathrm{W}/\sqrt{\mathrm{Hz}}],\\
    &=15~\mathrm{pW}/\sqrt{\mathrm{Hz}},\label{eq:NEP_val}
\end{align} this refers to the power required for a $\mathrm{SNR}=1$ as referred from the sensor intrinsic noise, for a detector efficiency $\eta_d=0.6$. Appendix~\ref{appendix:NEP} provides a derivation of $\mathrm{NEP}_\mathrm{sensor}$.

\section{SUPPORTING SYSTEM}
There are numerous supporting sub-systems required to obtain the desired atomic response. Each sub-system has its own, potentially contradictory, requirements which must be addressed. A static electric and magnetic field must be generated on the z-axis between the ends of the vapour cell to ensure the required Stark shift/mixing and quantisation-axis definition. The short length of the vapour cell loosens the voltage requirements for the DC field, but care must be taken to ensure that the incoming and generated fields can be accessed and be diverted away from the system. Moreover, the surrounding (non-dielectric) sub-systems will perturb the internal electric fields of the vapour cell: which must be accounted for.

Ensuring that the vapour cell is isolated from unwanted environmental effects is paramount. There should be enough shielding to reduce the impact of highly-energetic space radiation, which can induce ionisation and cause disruptive inhomogeneous electric fields. Likewise, there should be sufficient shielding from stray electric/magnetic fields which perturb the operating conditions. Stable heating and temperature control of the vapour cell is another concern, as the temperature of the system affects the BBR atomic transitions, vapour density, and various line-broadening effects.

The laser fields involved should have low phase noise; while we do not require phase-locking, the use of a common laser frequency reference would be highly beneficial. The optical/quasi-optical collection, collimation, transmission, and filtering needs care. For example, a signal at $4.7$~THz has a wavelength of $63$~$\mu$m, which places stringent manufacturing tolerances to avoid unwanted reflections and diffraction during propagation from the telescope to the vapour cell. The optical alignment of our fields must be robust, as the overlap of our fields is what defines the interaction region which underpins the operation of this system. From ionisation, ions can build up in the vapour cell, which can be highly disruptive to expected operation; hence, ionisation-clearing protocols could be investigated. There is also a concern pertaining to a small cell geometry, as surface charges from alkali deposition can perturb the system. 

For our photo-detector system our L-field detector should be moderately cooled to suppress dark-counts raising the noise floor. Further improvements to SNR can be achieved by spectral filtering before the photodetector, for example with a cavity \cite{Borowka24}. There can be a suite of other photo-detectors to monitor the atomic response: in particular, a P-field photo-detector can be employed to estimate the Rydberg atom population during a mode of operation for 3-level EIT.
\section{RADIOMETRY}
\subsection{Conventional THz radiometers}
Conventional THz radiometers are broadly divided into coherent heterodyne receivers and direct detectors \cite{Rogalsky22}. Heterodyne systems dominate high-resolution THz spectroscopy: SIS mixers provide very low noise below $(\sim1.2~\mathrm{THz})$, while superconducting hot-electron-bolometer (HEB) mixers extend operation to several THz. At $(4.7~\mathrm{THz})$, state-of-the-art HEB receivers achieve receiver noise temperatures of order $(800~\mathrm{K})$, corresponding to (radiometric sensitivity $\Delta T\sim0.8~\mathrm{K})$ for $(B=1~\mathrm{MHz})$ and $(t=1~\mathrm{s})$, but require cryogenic cooling and a stable supra-THz LO. Room-temperature Schottky mixers avoid cryogenics but are substantially noisier. Direct detectors such as TESs and KIDs can provide exceptional NEP, but generally require sub-kelvin cooling and additional spectroscopic optics to obtain MHz-scale frequency resolution. The principal challenges of conventional THz radiometry are therefore high receiver noise, difficult LO generation, cryogenic cooling, optical losses and calibration. 

A Rydberg THz-to-optical converter instead transfers the $(4.7~\mathrm{THz})$ signal into the optical domain, potentially removing both the cryogenic mixer and THz LO; its key challenge is then to achieve sufficiently high, stable and low-noise conversion efficiency. Care must be taken to ensure that the comparison between receiver technologies is fair \cite{Allinson26}, as the operating principles between these paradigms are considerably different.
\subsection{Calibration}
Calibration of the read-out to the incoming radiative power is a fundamental requirement of an operational radiometer. A common approach for calibration, for a radiometer with a linear relationship between incoming thermal power and output voltage over its dynamic range, is by the use of a hot and cold load. A blackbody emitter at a known temperature near the upper limit of the operational dynamic range, is known as the hot load. For a space radiometer, an unobstructed view to the dark sky can often function as the cold load, and provide the reference for the lower limit of the operational dynamic range. The linear power/voltage relationship enables interpolation of an unknown scene temperature against voltage. The stability of the hot (and cold) load is a concern for the long-term quality of this calibration technique. In particular, for higher frequencies such as the THz region, a challenge for this calibration technique is that the Rayleigh-Jeans approximation is invalid and Planck's law must be used. Planck's law is non-linear, hence error in measuring the hot/cold load temperature can translate to large calibration errors. Another concern is that for high-frequency radiometers, the small wavelength means that from microscopic surface imperfections, reflections and standing waves can form which disrupt the expected emissivity.

For calibration of our Rydberg THz radiometer concept, we can segment into three stages: scene to atom, the atomic response, and the atom to photodetection. Calibration of the scene to atom is similar to the challenges experienced by conventional THz radiometers: namely, unwanted reflections from surface imperfections. Calibration of the atomic response is unique to this concept compared to the other THz radiometer architectures. Any change in the response, due to for example, a varying Stark field, cell degradation, laser instability, must be monitored. The atomic response could be probed spectroscopically: a sequence of stable laser sources probing to constrain parameters pertaining to estimation of the atomic response. Furthermore, a known/well-characterised RF LO could also be used on the vapour cell to corroborate the RF LO field strength characterisation via the SI-traceable Autler-Townes (AT) splitting \cite{PhysRevLett.111.063001}. Successful corroboration of RF LO characterisation could indicate that our estimation of the atomic response is still valid. The two-way interconversion nature of the transducer could be leveraged \cite{Kiffner16}: a known L-field LO could be injected into the vapour cell, where absorption of the L-field is monitored. The inferred conversion of our L-field could assist in characterisation of the atomic response. The last stage, would likely be the easiest for calibration. There can be straightforward characterisation of the optical channel from cell to photo-detector. Lastly, a known L-field LO can be used to calibrate the photo-detector response.

\subsection{Radiometric figures of merit}
In our performance section, we assessed performance in terms of NEP. Whereas, many radiometers detail their performance through a noise temperature or radiometric sensitivity $\Delta T$. Care is required to ensure a fair comparison with conventional THz radiometers. The temperature variable must correspond to the same input brightness-temperature convention. Moreover, the accepted spatial/polarisation mode and optical throughput must be matched. 

We proceed with comparing our receiver concept with the performance of the Schottky-diode $4.7$~THz radiometer benchmark suggested by Hansen et al \cite{Hansen25}: one that is shown to support useful retrieval performance. Hence, we calculate the receiver-equivalent radiometric sensitivity, for integration time $\tau_d=1$~s and bandwidth $\mathcal{B}=1$~MHz as \begin{equation}
    \Delta T=\frac{\mathrm{NEP}_\mathrm{sensor}}{\sqrt{2\tau}k_B \mathcal{B}}=7.72\times 10^5~\mathrm{K},
\end{equation} using $\mathrm{NEP}_\mathrm{sensor}$ from Eq.~\ref{eq:NEP_val}. Our noise floor is a vacuum-reservoir lower bound as atomic fluorescence, technical noise and gain drift are not yet included. Accordingly, as our converted signal is weak, it is appropriate to describe the result as an idealised input-referred photon statistical $\Delta T$ for a fully filled accepted mode, rather than as a complete instrument-level $\Delta T$. 

The $4.7$~THz radiometer specified by Hansen et al. has a radiometric sensitivity $\Delta T=25$~K; clearly, our receiver design must be drastically improved to reach the demands of this benchmark. If optimisation of the detector efficiency $\eta_d$ and the $\mathrm{M}\to\mathrm{L}$ conversion-function $\vert t_{LM}\vert$ does not realise this demand, then a lower signal frequency could be targeted instead.

\subsection{Atomic oxygen atmospheric line emission}
As a case-example of sensing an atmospheric atomic species, we use Hansen et al.'s simulated spectral radiance incident at a receiver of the $4.7$~THz atomic oxygen emission \cite{Hansen25}. They solve a radiative transfer equation along a one-dimensional line of sight for a limb-sounding geometry, where the $4.7$~THz emission is assumed to be in local thermodynamic equilibrium and scattering is neglected. From the acquired brightness temperature profile $T_B(\omega)$ with FWHM $26$~MHz and using Planck's law, we calculate the normalised scene lineshape as \begin{equation}
    w(\omega)=\frac{\hbar\omega n_\mathrm{scene}(T_B(\omega))}{\int_\mathcal{B}\frac{d\omega^{'}}{2\pi}\hbar\omega^{'} n_\mathrm{scene}(T_B(\omega^{'}))}.
\end{equation} For a set coupling efficiency $\eta_\mathrm{c}=0.6$, from Eq.~\ref{eq:NEP_scene} we can calculate the atomic oxygen scene-referred NEP as $\mathrm{NEP}_\mathrm{scene}=112~\mathrm{pW}/\sqrt{\mathrm{Hz}}$. This quantifies the scene input-referred noise for a perturbation having the spectral shape of the atomic-oxygen emission line. The incident atomic oxygen emission line power within the accepted mode and receiver bandwidth is $P_\mathrm{OI}=\int_\mathrm{B}\frac{d\omega}{2\pi}\hbar\omega n_\mathrm{scene}(\omega)=99.47$~fW. Consequently, we have a SNR$=\frac{\sqrt{2\tau}P_\mathrm{OI}}{\mathrm{NEP}_\mathrm{scene}}=1.88\times 10^{-3}$. The simulated data we used for the atomic oxygen spectra corresponds to the parameters: $\tau=2.26$~s, tangent height $h=147.28$~km, scan number 14, and date ``2022-09-07T10:40:11".

The present receiver exhibits a poor signal-to-noise ratio for the atomic oxygen science case, arising not only from the weak absolute conversion efficiency but also from a substantial mismatch between the spectral conversion response $(\vert t_{LM}(\omega)\vert^2$ and the approximately Gaussian $4.7~\mathrm{THz}$ atomic oxygen emission profile. This motivates treating the driven atomic medium as a tunable spectral pre-filter, with the control-field detunings, Rabi frequencies, and residual Doppler broadening tailored to increase the overlap between the conversion response and the lineshape of the atomic oxygen signal. Exact matching of the atomic response to the emission profile is not necessarily optimal, since broadening the response may reduce peak conversion and admit additional broadband noise; the relevant design objective is therefore to minimise the atomic oxygen-referred NEP, with spectral overlap used as a diagnostic. If sufficient conversion and suitable spectral selectivity can be achieved, such a line-integrating radiometer could provide a compact measurement of the receiver-weighted atomic oxygen line power. With an independent co-located profile of atmospheric temperature, this line-integrated observable can then be related through a limb radiative-transfer model to the atomic oxygen abundance along the line of sight.

\section{DISCUSSION}
This paper has set out to provide an assessment of the feasibility of a THz line-selective radiometer which does not need cryogenic cooling systems or a LO at the frequency band of the signal of interest. A theoretical model has been built of a lossy coherent transducer of a quantised thermal input THz field to an output quantised optical field. The output photon statistics have been calculated, with various noise sources setting a noise floor. System performance is then derived from the photon statistics. A discussion on the system requirements is given and the concept is contextualised to radiometry, whereby a case-example of sensing the $4.7$~THz atomic oxygen emission line is detailed.
 
The fidelity of this model can be substantially improved by the inclusion of the stochastic and inhomogeneous electric field generated by ionisation events and a full (non-collinear) treatment of the field wave-vectors. The fidelity can also be improved by considering prominent technical noise sources that perturb the system: relative laser amplitude noise, vapour cell temperature variation, and DC-field instabilities. Accurate accounting of fluorescence generated at the L-field frequency should also be included in future models of this system.

System performance can be improved via optimisation of the atomic response from the model presented. This optimisation could involve the development of a Fisher information framework similar to recent work for an EIT-AT electrometer \cite{Rong26}. The selected manifold of states, vapour cell temperature, Stark tuning, control-field Rabi frequencies, and control-field detunings constitute the main variables for optimisation. Furthermore, appropriate tailoring of the atomic response could realise this system as a matched pre-filter for the signal of interest: reducing out-of-band perturbations and reducing unnecessarily coupled sensor-induced BBR. From a systems engineering aspect, there are many challenges for a Rydberg conversion-based sensor to proceed towards being a feasible space radiometer. The inherent system complexity is apparent due to the number of stable lasers required. This complexity is compounded by the requirement of a well-shielded, stable temperature, electric and magnetic field environment for the vapour cell. There is also the added complexity pertaining to ionisation-clearing/accounting and calibration protocols. It is clear that improved fidelity models and adequate operation routines (calibration and maintenance) are still to be developed, prior to system integration and sub-system maturity. 

It has been experimentally demonstrated for a Rydberg conversion-based system that input field phase modulation is preserved for the transduced signal \cite{Han18}. As such further developments could involve optical heterodyning of the thermal transduced signal, which would avoid the use of a potentially difficult LO at the input field signal frequency. However, noise from the (back-end) optical LO may represent a significant barrier to this development. If such a development is possible, the resultant fine-resolution could facilitate detection of neutral winds from the Doppler shift in the atomic oxygen emission. It could also facilitate detection of atomic oxygen at relatively low altitudes, as the lineshape wings could be analysed. A high-fidelity model would be required for the kHz spectral resolution. Lastly, the vapour cell could be placed within an M-field resonant cavity, provided that it does not compromise laser access or extraction of the generated field. This would enable further spectral filtering and could provide local field enhancement to improve the M-field to atom coupling \cite{Liu25}. To conclude, maturation of the sub-system technology and optimisation of system performance with a high-fidelity model could pave the way to a competitive Rydberg conversion-based THz radiometer.

\appendix
\section{Photon-flux normalisation}
\label{appendix:normalisation}
We wish to demonstrate how our travelling continuum mode operator is normalised to photon-flux. The Fourier transform of the frequency travelling-mode operator (narrow-band approximation) $\hat{a}_x(\omega)$ is \begin{equation}
    \hat{a}_x(t)=\int^\infty_0 \frac{d\omega}{2\pi}\hat{a}_x(\omega)e^{-i\omega t},
\end{equation} which, for the narrowband approximation, yields the commutator $[\hat{a}_x(t),\hat{a}_x(t^{'}]\approx\delta(t-t^{'})$, therefore the operator $\hat{a}^\dag_x(t)\hat{a}_x(t)$ is interpreted as a photon-flux operator with units photons/s. The total photon flux is written as \begin{align}
    \Phi_x&=\langle \hat{a}_x^\dag(t)\hat{a}_x(t)\rangle,\\ 
    &=\int^\infty_0 \frac{d\omega}{2\pi}n_x(\omega),
\end{align} thus the photon-flux spectral density is $\frac{d\Phi}{d\omega}=\frac{n_x(\omega)}{2\pi}$. The field envelope was selected to ensure that the power spectral density corresponds to the photon-flux spectral density multiplied by $\hbar\omega$, 
\section{Transfer function derivation}\label{appendix:transfer_function}
In this appendix we derive the solution of the nonlinear wave equation Eq.~\ref{eq:NL_waveeqn}. For brevity we define the following expressions \begin{align*}
    A(\omega)&=\eta_M(\omega) \xi^M_{43}(\omega), \\
    B(\omega)&=\eta_M(\omega) \xi^L_{43}(\omega), \\
    C(\omega)&=\eta_L(\omega_L)\xi^M_{61}(\omega), \\
    D(\omega)&=\eta_L(\omega_L)\xi^L_{61}(\omega), \\
    \mu(\omega)&=\frac{1}{2}(A(\omega)+D(\omega)),\\
    K(\omega)&=\sqrt{\left(\frac{A(\omega)-D(\omega)}{2}\right)^2+B(\omega)C(\omega)},
\end{align*} where the coupling coefficients $\eta_M(\omega)$ and $\eta_L(\omega_L)$ are \begin{align}
    \eta_M(\omega)&=\mathcal{N}\frac{\vert d_{43}\vert^2}{2\hbar \epsilon_0 c}\omega, \\
    \eta_L(\omega_L)&=\mathcal{N}\frac{\vert d_{61}\vert^2}{2\hbar \epsilon_0 c}\omega_L,
\end{align} the $\omega$ argument describes a spectral component of the M-field and the L-field's frequency $\omega_L$ is linearly dependent on $\omega$, due to loop-closure conditions.

In the paraxial-wave approximation, ignoring diffraction, invoking the slowly-varying wave-envelope approximation, and transforming to a frame $(t,z) \to (\tau=t-z/c,z)$ co-moving with the M and L-fields we have a coupled set of differential equations derived from Eq.~\ref{eq:NL_waveeqn} expressed in vectorial form \begin{equation}
    \frac{\partial}{\partial z}\begin{pmatrix}
        \Omega_M(z,\omega) \\ \Omega_L(z,\omega_L)
    \end{pmatrix}= i\begin{pmatrix}
        A(\omega) & B(\omega) \\ C(\omega) & D(\omega) \label{eq:differential}
    \end{pmatrix}\begin{pmatrix}
        \Omega_M(z,\omega) \\ \Omega_L(z,\omega_L)
    \end{pmatrix},
\end{equation} The independence of spectral components in Eq.~\ref{eq:differential} is valid, granted that we are in an adiabatic, linear and stationary regime. The solution is hence \begin{equation}
    \begin{pmatrix}
        \Omega_M(z,\omega) \\ \Omega_L(z,\omega_L)
    \end{pmatrix} = H(z,\omega) \begin{pmatrix}
        \Omega_M(0,\omega) \\ \Omega_L(0,\omega_L)
    \end{pmatrix}=\begin{pmatrix}
       H_{MM} & H_{ML} \\ H_{LM} & H_{LL}
    \end{pmatrix}\begin{pmatrix}
        \Omega_M(0,\omega) \\ \Omega_L(0,\omega_L)
    \end{pmatrix},\label{eq:input_output_rabi}
\end{equation} where the transfer function $H(\omega,z)$ is defined as
\begin{equation}
H(\omega,z)=e^{\frac{i}{2}(A(\omega)+D(\omega))z}\begin{pmatrix}
        \mathrm{cos}(K(\omega)z)+i\frac{A(\omega)-D(\omega)}{2K(\omega)}\mathrm{sin}(Kz) & i\frac{B(\omega)}{K(\omega)}\mathrm{sin}(K(\omega)z) \\ i\frac{C(\omega)}{K(\omega)}\mathrm{sin}(K(\omega)z) & \mathrm{cos}(K(\omega)z)-i\frac{A(\omega)-D(\omega)}{2K(\omega)}\mathrm{sin}(K(\omega)z)
    \end{pmatrix}.
    \end{equation}

\section{NEP and radiometric sensitivity derivation}
A small change of incident power in the accepted mode is \begin{equation}
    \delta P=\int_\mathcal{B}\frac{d\omega}{2\pi}\hbar\omega \delta n(\omega),
\end{equation} where $\delta n(\omega)=\frac{\delta P}{\hbar\omega}w(\omega)$ and we have introduced a bandwidth-normalised lineshape $w(\omega)$. To characterise the intrinsic receiver noise, we set $n_\mathrm{scene}(\omega)=0$ and refer the remaining noise to an equivalent power perturbation at the input of the atomic converter. Thus, the corresponding change of mean photo-count is \begin{equation}
    \delta\langle N_L(n_\mathrm{scene}(\omega)=0)\rangle=\tau_d\int_\mathcal{B}\frac{d\omega}{2\pi\hbar\omega}\eta_d\vert t_{LM}\vert^2\delta P_\mathrm{sensor} w(\omega).
\end{equation} For a perturbation referred instead to the incident scene, input-field coupling efficiency must be included,  \begin{equation}
    \delta\langle N_L\rangle=\tau_d\int_\mathcal{B}\frac{d\omega}{2\pi\hbar\omega}\eta_d\eta_\mathrm{c}\vert t_{LM}\vert^2\delta P_\mathrm{scene} w(\omega).
\end{equation}

For a signal-to-noise ratio of \begin{equation}
    \mathrm{SNR}=\frac{\delta \langle N_L\rangle}{\sqrt{\mathrm{Var}(N_L)}},
\end{equation} we set $\mathrm{SNR}=1$ and solve for the minimum power required $\delta P_\mathrm{min}$. Hence, the intrinsic input-referred noise equivalent power (NEP) is defined as \begin{equation}
\mathrm{NEP}_\mathrm{sensor}=\frac{\delta P_\mathrm{sensor,min}}{\sqrt{\Delta f_\mathrm{meas}}}=\frac{\sqrt{2\tau_d\mathrm{Var}(N_L(n_\mathrm{scene}(\omega)=0))}}{\vert\frac{\partial\langle N_L(n_\mathrm{scene}(\omega)=0)\rangle}{\partial P_\mathrm{sensor}}\vert}=\frac{\sqrt{2\tau_d\mathrm{Var}(N_L(n_\mathrm{scene}(\omega)=0))}}{\tau_d\int_\mathcal{B}\frac{d\omega}{2\pi\hbar\omega}\eta_d\vert t_{LM}\vert^2 w(\omega)} \;[\mathrm{W}/\sqrt{\mathrm{Hz}}],
\end{equation} using the one-sided measurement equivalent noise bandwidth convention $\Delta f_\mathrm{meas}=\frac{1}{2\tau_d}$, where we use a top-hat function for $w(\omega)$ to compare receiver performance. Whereas, for the scene input-referred noise equivalent power we have \begin{equation}
\mathrm{NEP}_\mathrm{scene}=\frac{\delta P_\mathrm{scene,min}}{\sqrt{\Delta f_\mathrm{meas}}}=\frac{\sqrt{2\tau_d\mathrm{Var}(N_L)}}{\vert\frac{\partial\langle N_L\rangle}{\partial P_\mathrm{scene}}\vert}=\frac{\sqrt{2\tau_d\mathrm{Var}(N_L)}}{\tau_d\int_\mathcal{B}\frac{d\omega}{2\pi\hbar\omega}\eta_d\eta_\mathrm{c}\vert t_{LM}\vert^2 w(\omega)} \;[\mathrm{W}/\sqrt{\mathrm{Hz}}],\label{eq:NEP_scene}
\end{equation} where we use $w(\omega)$ to correspond to the signal of interest lineshape, which for our case example is atomic oxygen. 

\label{appendix:NEP}
\acknowledgements{The authors would like to thank Aaron Strangfeld for the useful discussions.}
\bibliography{THz_bib} 
\bibliographystyle{spiebib} 
\end{document}